\documentclass[12pt]{iopart}
\usepackage{graphicx}

\def\bea{\begin{eqnarray}}
\def\eea{\end{eqnarray}}
\newcommand{\deta}{\ensuremath{\eta_\Delta}}
\newcommand{\dphi}{\ensuremath{\phi_\Delta}}
\newcommand{\drsref}{\ensuremath{\frac{\Delta\rho}{\sqrt{\rho_{\rm ref}}}}}

\begin{document}
  
%\article[Short Title]{Heading}{Title} 
\article[Minijet Correlations]{Quark Matter 2008}
	{Anomalous centrality variation of minijet angular correlations 
	  in Au-Au collisions at 62 and 200 GeV from STAR}
\author{M Daugherity for the STAR Collaboration}
\address{University of Texas, Department of Physics, Austin TX 78712, USA}
\ead{daugherity@physics.utexas.edu}

\begin{abstract}
  We have measured 2D autocorrelations for all
  charged hadrons in STAR with $p_{t}$ $>$ 0.15 GeV/c and $| \eta |$  $<$ 1 from
  Au+Au collisions at 62 and 200 GeV. The correlation structure is
  dominated by a peak centered at zero relative opening angles on $\eta$
  and $\phi$ which we hypothesize is caused by minimum-bias jets
  (minijets). We observe a large excess of minijet correlations
  in more-central Au-Au collisions relative to binary-collision scaling
  (more correlated pairs than expected from surface emission or even
  volume emission). We also
  observe a sudden increase of the minijet peak amplitude
  and $\eta$ width relative to binary-collision scaling of scattered partons which occurs at an
  energy-dependent centrality point. 
  There is a possible scaling of the transition point with
  transverse particle density. 
  %The large increase of the minijet correlations from peripheral to
  %central Au-Au collisions appears to be strongly inconsistent with
  %thermalization in heavy ion collisions.
  The large minijet correlations bring into question the 
  degree of thermalization in RHIC collisions.

\end{abstract}
%\pacs{} %% CHECK THESE LATER
%\submitto{\jpg}  %% ?
%\maketitle

%%%%%%%%%%%%%%%%%%%%%%%%%%%%%%%%%%%%%%%%%%%%%%%%%%%%%%%%%%%%%%%%%
\section{Introduction}

% Provide a quick motivation for studying minijets
Low momentum jets are estimated to produce 50\% of transverse energy in RHIC heavy ion
collisions and 80\% at the LHC \cite{minijet}.
Despite the large role these \emph{minijets} play, they have received little 
attention from the general community since the start of the RHIC experimental program.
As minijet abundance increases
at higher energies, the dynamics of minijet interactions are becoming essential to
understanding heavy ion collisions.   

% Minbias correlations, minijet definition, overview
%While low momentum jets are not individually resolvable, their aggregate effect generates an observable
%two-particle correlation.  In a theoretical context, minijets are typically defined 
%within the range of applicability of pQCD by specifying a low hadron $p_t$ cutoff around 2 GeV/c, 
%even though QCD interactions continue to lower momenta.  
%Therefore, we will experimentally
%define minijets based on correlation structure rather than an {\it a priori} $p_t$ range.  
%This approach requires a \emph{minimum-bias} two-particle correlation analysis
%where every possible pair of particles is considered instead of selecting a few trigger/associated
%pairs. Minijets are distinguished from other sources by decomposing the unique correlation structures.  
%Previous analyses have used this technique to reveal large minijet contributions
%in transverse \cite{130xtxt} and axial ($\eta$, $\phi$) \cite{130CI} spaces at 130 GeV 
%at four centralities. Here we report
%the detailed energy and centrality dependence of minijet angular correlations at RHIC. 

While low momentum jets are not individually resolvable, their combined effect generates an observable
correlation.  In a theoretical context minijets are typically defined 
within the range of applicability of pQCD by specifying a low hadron $p_t$ cutoff around 2 GeV/c, 
even though QCD interactions continue to lower $p_t$.  
We experimentally
define minijets based on correlation structure rather than an {\it a priori} $p_t$ range.  
This requires a \emph{minimum-bias} two-particle correlation analysis
where every possible pair of particles is considered instead of selecting a few trigger/associated
pairs. Minijets are distinguished from other sources by decomposing the unique correlations.  
Previous analyses have used this technique to reveal large minijet contributions
in transverse \cite{130xtxt} and axial ($\eta$, $\phi$) \cite{130CI} spaces at 130 GeV 
at four centralities. Here we report
the detailed energy and centrality dependence of minijet angular correlations at RHIC. 

%%%%%%%%%%%%%%%%%%%%%%%%%%%%%%%%%%%%%%%%%%%%%%%%%%%%%%%%%%%%%%%%%
\section{Analysis}

% A few details on data, correlation measure

\begin{figure}[htb]
  \begin{center}
    \includegraphics[width=\textwidth]{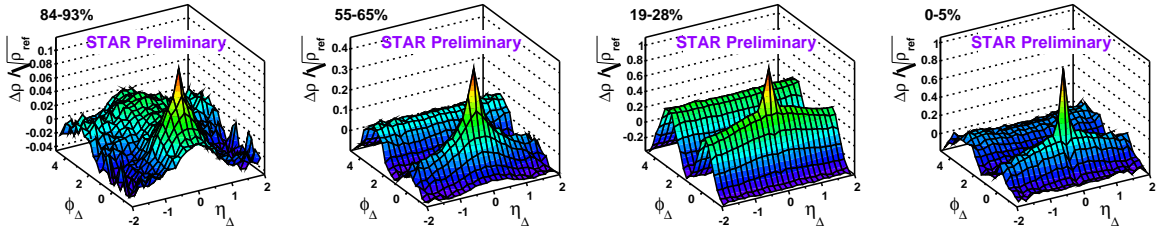}
    \caption{Minimum-bias correlations for several centralities from peripheral (left) 
      to central (right) in 200 GeV Au+Au collisions.}
    \label{fig1}
  \end{center}
\end{figure}

Charged particle tracks detected in the STAR TPC with $p_t > 0.15$ GeV/c, 
$|\eta| < 1$, and full $2\pi$ azimuth were analyzed from 1.2M minbias triggered 
200 GeV Au+Au and 6.7M 62 GeV Au+Au events.  
%Pair densities $\rho(\vec{p}_1, \vec{p}_2)$ 
%measured as number of pairs per unit area on
%were measured for all possible unique particle pairs
%as a function of relative angles $\deta \equiv \eta_1 - \eta_2$ and $\dphi \equiv \phi_1 - \phi_2$.
Pair densities $\rho(\vec{p}_1, \vec{p}_2)$ were measured as number of pairs per unit area on
relative angles ($\deta \equiv \eta_1 - \eta_2$, $\dphi \equiv \phi_1 - \phi_2$) 
for all possible unique particle pairs.
Particles within the same event form \emph{sibling} pair densities $\rho_{sib}$, 
while mixing particles from different events measures the uncorrelated \emph{reference} $\rho_{ref}$.  
%These densities are formed into a correlation measure by applying  
%the standard definition of a correlation as a normalized covariance.
These are formed into a normalized covariance to produce a correlation measure.
The difference 
$\Delta \rho \equiv \rho_{sib} - \rho_{ref}$ measures the covariance in number of pairs 
between histogram bins, and the normalization is provided by bin-wise division of  
$\sqrt{\rho_{ref}}$.  Thus we use the notation \drsref{} for a \emph{per-particle} 
correlation measure, shown in figure \ref{fig1} for selected centralities.
 
%%%%%%%%%%%%%%%%%%%%%%%%%%%%%%%%%%%%%%%%%%%%%%%%%%%%%%%%%%%%%%%%%
\section{Fit Results}

\begin{figure}[htb]
  \begin{center}
    \includegraphics[width=\textwidth]{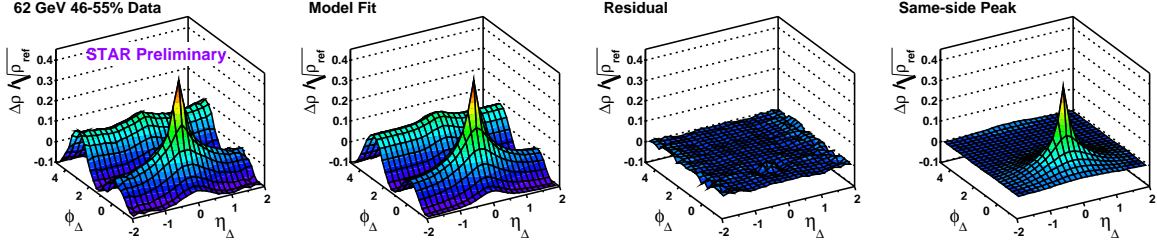}
    \caption{An example of the fit function showing   
      correlation data (first panel), model function (second panel), 
      residual (third panel) defined as data minus model fit,
      and the same-side gaussian and exponential peaks (last panel).}
    \label{fig2}
  \end{center}
\end{figure}

% pp components, fit function
Proton-proton collisions provide a reference for measuring the contributions to these  
structures.  Analysis of minimum-bias correlations \cite{ppcorr} and single particle $p_t$ 
spectra \cite{ppspectra} show that p+p collisions are well described by a two-component 
soft and semi-hard scattering model, as commonly used in event generators such as 
\textsc{Pythia}.  %\cite{Pythia}.  
The soft component represents longitudinal fragmentation
in unlike-sign pairs and produces a 1D gaussian correlation centered along \deta=0.  
The semi-hard component contains a same-side peak, modeled as a 2D gaussian at the 
$\deta=\dphi=0$ origin,
%and an away-side peak centered at azimuth difference \dphi=$\pi$.  
and an away-side ridge centered at \dphi=$\pi$.  
For an inclusive $p_t$ range the away-side is completely represented by function $-\cos(\dphi)$
that approximates a wide gaussian which narrows with increasing $p_t$ \cite{ppcorr}.
The final
component necessary to describe p+p data is a 2D exponential at the origin containing 
contributions from HBT in like-sign pairs and conversion $e^{\pm}$ in unlike-sign pairs.  
%To ensure a well physically motivated fit function for Au+Au collisions, 
To ensure the simplest possible fit function for Au+Au collisions, 
we use these
components from p+p collisions with only one additional $\cos(2\dphi)$ quadrupole term
to account for
correlations conventionally attributed to elliptic flow \cite{FlowMethods}.  The eleven
parameter fit function used for the correlation structures in figure \ref{fig1} is then: \\ 
\bea \label{eq:fit}
%F &  =&  A_{\phi_\Delta}\, \cos(\phi_\Delta) + A_{2\phi_\Delta}\, \cos(2\, \phi_\Delta) \nonumber \\
%&&+ A_0\, \exp\left[- \left( \frac{\eta_{\Delta}}{\sqrt{2} \sigma_{0}} \right)^2 \right] 
%+ A_1 \, \exp \left\{ - \left[ \left( \frac{\phi_{\Delta}}{\sqrt{2} \sigma_{\phi_{\Delta}}} \right)^2  + \left( \frac{\eta_{\Delta}}{\sqrt{2} \sigma_{\eta_{\Delta}}} \right)^2  \right] \right\}  \nonumber \\
% &&+ A_2 \, \exp \left\{ - \left[ ( \phi_{\Delta}/w_{\phi_{\Delta}} )^2  + (\eta_{\Delta}/ w_{\eta_{\Delta}} )^2 \right]^{1/2} \right\}  + A_3.
%F &=&  A_{\phi_\Delta}\cos(\phi_\Delta) + A_{2\phi_\Delta}\cos(2\, \phi_\Delta) 
%+ G_s(\deta: A_0, \sigma_0) \nonumber \\
% && + G_h(\deta, \dphi: A_1, \sigma_{\eta\Delta}, \sigma_{\phi\Delta})
%+ E(\deta, \dphi: A_2, w_{\eta\Delta}, w_{\phi\Delta}) + A_3
\fl F =  A_{\phi_\Delta}\cos(\phi_\Delta) + A_{2\phi_\Delta}\cos(2\, \phi_\Delta) 
+ G_s(\deta: A_0, \sigma_0) 
+ G_h(\deta, \dphi: A_1, \sigma_{\eta\Delta}, \sigma_{\phi\Delta}) \nonumber \\
+ E(\deta, \dphi: A_2, w_{\eta\Delta}, w_{\phi\Delta}) + A_3
%+ A_0\,G_s(\deta: \sigma_0) \nonumber \\
% && + A_1\,G_h(\deta, \dphi: \sigma_{\eta\Delta}, \sigma_{\phi\Delta})
%+ A_2\,E(\deta, \dphi: w_{\eta\Delta}, w_{\phi\Delta}) + A_3
\eea
where $G_s$ and $G_h$ are the soft and hard Gaussian terms and $E$ is an exponential
function with parameters listed after the colon.
An example of this fit is shown in figure \ref{fig2}.

\begin{figure}[htb]
  \begin{center}
    \includegraphics[width=\textwidth]{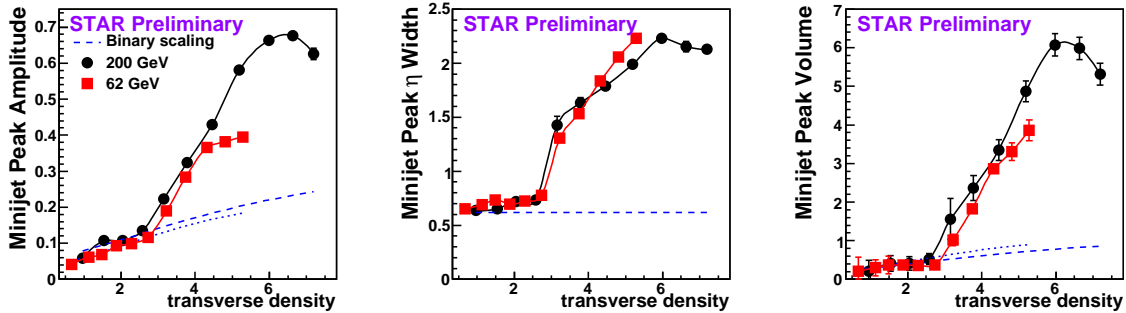}
    \caption{Same-side gaussian peak amplitude, \deta{} width, and volume fits.
      Points show eleven centrality bins for each energy 
      (84-93\%, 74-84\%, 65-74\%, 55-65\%, 46-55\%, 37-46\%, 28-37\%, 19-28\%, 9-19\%, 5-9\%, and 0-5\%)
      transformed to tranvserse density.}
    \label{fig3}
  \end{center}
\end{figure}

% fig 3, binary scaling, transition, transverse density, 
Figure \ref{fig3} shows the measured fit parameters for the same-side peak amplitude, 
\deta{} width, and volume ($=2\pi A_1 \sigma_{\eta\Delta} \sigma_{\phi\Delta}$).
Fitting errors are shown and systematic error 
is estimated to be $\pm$9\% of the correlation amplitude
and at most a few percent of the widths.      
The dashed lines show the binary scaling reference expected from independent nucleon-nucleon 
collisions.  Using the Kharzeev and Nardi two-component model \cite{KN} and path length
$\nu \equiv 2\langle N_{bin} \rangle / \langle N_{part} \rangle$, the minijet amplitude 
in Au+Au collisions is expected
to scale as $A_1(\nu) = A_{1,pp} \: \nu / [1 + x(\nu - 1)]$ from the p+p value.
Peripheral collisions follow the binary scaling reference closely, deviating only by small
increases in \deta{} and decreases in \dphi{} widths.  The data show a sharp
transition at approximately 55\% centrality for 200 GeV and 40\% for 62 GeV where the
amplitude and \deta{} widths increase dramatically while the \dphi{} widths continue
to decrease slightly.  
Centrality in figure \ref{fig3} is represented by transverse particle density calculated as 
$\frac{3}{2} \frac{dN_{ch}}{d\eta} / \langle S \rangle$ 
with initial collision overlap area $\langle S \rangle$ from Monte Carlo Glauber.
%Using this quantity gives excellent agreement between the two energies, 
Transverse density brings the  transition points for the two energies to coincidence,
whereas conventional centrality measures displace the transition points
and tend to compress the peripheral data.  

%%%%%%%%%%%%%%%%%%%%%%%%%%%%%%%%%%%%%%%%%%%%%%%%%%%%%%%%%%%%%%%%%
\section{Discussion}

The correlation structures are modified at the transition,
but are still likely to be associated with minijets for several reasons.  First, these
results, particularly when taken with a similar analysis of $p_t$ correlations \cite{ptcorr}, 
show that contributions from a new physical mechanism unrelated to minijets are unlikely.
Any such hypothetical process must have \dphi{} widths and 
$p_t$ correlations that match seamlessly with
minijets, which would be a remarkable coincidence.  
Second, the amplitude and \deta{} width increases are consistent with further minijet interactions,
which may be possible due to path-length considerations \cite{MinijetProduction}.
Finally, it is possible that the new correlation structures are due to changes in 
minijet fragmentation.  
The trends in the data also suggest a lower $p_t$ manifestation of the ``ridge'' \cite{Ridge},
and these results may help to discriminate among the many competing models of ridge formation.

% Yield
The same-side peak volume gives the total number of correlated pairs, though finding
the particle yield requires estimating the average number of correlated structures
per event. Assuming each structure originates with a semi-hard parton and that 
semi-hard scattering follows binary scaling,
we estimate that 30\% of all final-state hadrons
in central 200 GeV Au+Au collisions are associated with this same-side correlation.  

% Thermalization
As a source of correlated low momentum particles, minijets provide an extremely
sensitive probe of the collision system.
The binary scaling reference represents one extreme limit of a transparent medium,
while the other extreme is a completely thermalized system opaque 
to minijets \cite{Therm}.  
These results call into question the existence of the latter system at RHIC energies.
%These results call into question whether or not the correlation among partons 
%from minijet production in the early stages of the collision would survive to the final-state
%hadrons in such a system.

%%%%%%%%%%%%%%%%%%%%%%%%%%%%%%%%%%%%%%%%%%%%%%%%%%%%%%%%%%%%%%%%%


\Bibliography{10}

\bibitem{minijet} Wang X-N and Gyulassy M 1991 {\it Phys. Rev. D} {\bf 44} 3501
\bibitem{130xtxt} Adams J \etal (STAR Collaboration) 2007 {\it J. Phys. G: Nucl. Part. Phys.} {\bf 34} 799
\bibitem{130CI} Adams J \etal (STAR Collaboration) 2006 {\it Phys. Rev. C} {\bf 73} 064907
\bibitem{ppcorr} Porter R J and Trainor T A 2005 {\it J. Phys.: Conf. Ser.} {\bf 27} 98
\bibitem{ppspectra} Adams J \etal (STAR Collaboration) 2006 {\it Phys. Rev D} {\bf 74} 032006
%\bibitem{Pythia} Sj\"{o}strand T \etal 2006 {\it J. High Energy Phys.}{\bf 05} 026
\bibitem{FlowMethods} Trainor T A and Kettler D T 2007 {\it Preprint} arXiv:0704.1674 [hep-ph]
\bibitem{KN} Kharzeev D and Nardi M 2001 {\it Phys. Lett. B} {\bf 507} 121
\bibitem{ptcorr} Adams J \etal (STAR Collaboration) 2006 {\it J. Phys. G: Nucl. Part. Phys.} {\bf 32} L37
\bibitem{MinijetProduction} Kajantie K \etal 1987 {\it Phys. Rev. Lett.} {\bf 59} 2527 
\bibitem{Ridge} Putschke J (STAR Collaboration) 2007 {\it Preprint} nucl-ex/0701074
\bibitem{Therm} Nayak G C \etal 2001 {\it Nucl. Phys. A} {\bf 687} 457;  
  Shin G R and M\"{u}ller 2003 {\it J. Phys. G: Nucl. Part. Phys.} {\bf 29} 2485
\endbib
\end{document}